\documentclass[fleqn,usenatbib]{mnras}

\usepackage{newtxtext,newtxmath}
\usepackage[T1]{fontenc}

\DeclareRobustCommand{\VAN}[3]{#2}
\let\VANthebibliography\thebibliography
\def\thebibliography{\DeclareRobustCommand{\VAN}[3]{##3}\VANthebibliography}

\usepackage{graphicx}	% Including figure files
\usepackage{amsmath}	% Advanced maths commands
\newcommand{\sect}[1]{Section\,\ref{#1}}

\newcommand{\fig}[1]{Figure\,\ref{#1}}
\newcommand{\figs}[1]{Figures\,\ref{#1}}
\newcommand{\tab}[1]{Table\,\ref{#1}}
\newcommand{\tabs}[1]{Tables\,\ref{#1}}
\newcommand{\eqn}[1]{Equation\,(\ref{#1})}
\newcommand{\eqns}[1]{Equations\,(\ref{#1})}

\title[MIT for EIS]{Solar coronal magnetic field measurements using spectral lines available in \textit{Hinode}/EIS observations: Strong and weak field techniques and temperature diagnostics}

\author[Chen et al.]{
Yajie Chen,$^{1,2}$
Xianyong Bai,$^{3}$\thanks{E-mail: xybai@bao.ac.cn}
Hui Tian,$^{2,4}$
Wenxian Li,$^{3}$
Feng Chen,$^{5,6}$
Zihao Yang,$^{2}$
Yang Yang$^{7,8}$
\\
$^1$Max Planck Institute for Solar System Research, G\"{o}ttingen 37077, Germany\\
$^{2}$School of Earth and Space Sciences, Peking University, Beijing 100871, People's Republic of China\\
$^{3}$National Astronomical Observatories, Chinese Academy of Sciences, Beijing 100012, People's Republic of China\\
$^{4}$Key Laboratory of Solar Activity and Space Weather, National Space Science Center, Chinese Academy of Sciences, Beijing 100190, People's Republic of China\\
$^{5}$School of Astronomy and Space Science, Nanjing University, Nanjing 210023, People's Republic of China\\
$^{6}$Key Laboratory for Modern Astronomy and Astrophysics (Nanjing University), Ministry of Education, Nanjing 210023, People's Republic of China\\
$^{7}$Shanghai EBIT laboratory, Institute of Modern Physics, Fudan University, Shanghai, People's Republic of China\\
$^{8}$Key Laboratory of Nuclear Physics and Ion-beam Application (MOE), Fudan University, Shanghai 200433, People's Republic of China
}

\date{Accepted XXX. Received YYY; in original form ZZZ}

\pubyear{2022}

\begin{document}
\label{firstpage}
\pagerange{\pageref{firstpage}--\pageref{lastpage}}
\maketitle

% Abstract of the paper
\begin{abstract}
Recently, it has been proposed that the magnetic-field-induced transition (MIT) in Fe~{\sc{x}} can be used to measure coronal magnetic field strengths. 
Several techniques, the direct line ratio technique and the weak and strong magnetic field techniques, are developed to apply the MIT theory to spectroscopic observations taken by EUV Imaging Spectrometer (EIS) onboard \textit{Hinode}.
However, the suitability of coronal magnetic field measurements based on the weak and strong magnetic field techniques has not been evaluated.
Besides, temperature diagnostics is also important for measuring coronal magnetic field based on the MIT theory, but how to determine the accurate formation temperature of the Fe~{\sc{x}} lines from EIS observations still needs investigation.
In this study, we synthesized emissions of several spectral lines from a 3D radiation magnetohydrodynamic model of a solar active region, and then derived magnetic field strengths using different methods.
We first compared the magnetic field strengths derived from the weak and strong magnetic field techniques to the values in the model.
Our study suggests that both weak and strong magnetic field techniques underestimate the coronal magnetic field strength.
Then we developed two methods to calculate the formation temperature of the Fe~{\sc{x}} lines.
One is based on differential emission measure analyses, and the other is deriving temperature from the Fe~{\sc{ix}} and Fe~{\sc{xi}} line pairs. 
However, neither of the two methods can provide temperature determination for accurate coronal magnetic field measurements as those derived from the Fe~{\sc{x}} 174/175 and 184/345 {\AA} line ratios.
More efforts are still needed for accurate coronal magnetic field measurements using EIS observations.

\end{abstract}

% Select between one and six entries from the list of approved keywords.
% Don't make up new ones.

%%%%%%%%%%%%%%%%%%%%%%%%%%%%%%%%%%%%%%%%%%%%%%%%%%

%%%%%%%%%%%%%%%%% BODY OF PAPER %%%%%%%%%%%%%%%%%%

%==========================================================================================
\section{Introduction}
%==========================================================================================

Most activities in the solar corona are governed by the magnetic field, and routine and accurate coronal magnetic field measurements are the keys to understanding the dynamics in the corona.
However, the coronal magnetic field measurements are limited \citep[e.g.,][]{Wiegelmann2014}.

The Zeeman effect has been widely applied to Stokes profiles of spectral lines to achieve magnetic field measurements in the photosphere \citep[e.g.,][]{Iniesta2016,2019LRSP...16....1B} and chromosphere \citep[e.g.,][]{NICOLE,STiC}.
Nevertheless, it is challenging to measure the coronal magnetic field based on the Zeeman effect.
\citet{Lin2000,Lin2004} endeavored to measure the coronal magnetic field in an off-limb active region from spectropolarimetric observations of the infrared Fe~{\sc{xiii}} 10747 {\AA} line.
But the data was integrated over 70 minutes to obtain a high signal-to-noise ratio, impeding the investigation of the temporal evolution of the coronal magnetic field.
Besides, these measurements are limited to the observations above the limb.

Radio observations have been used to diagnose coronal magnetic fields in active regions \citep[e.g.,][]{Akhmedov1982,Akhmedov1986,Wang2015,Miyawaki2016,Iwai2013,Anfinogentov2019} and flaring structures \citep[e.g.,][]{Gary2018,Chen2020,Fleishman2020,Zhu2021} based on spectra fitting.
The diagnostic functions of the magnetic field highly rely on the emission mechanisms in radio observations. However, the determination of the emission mechanisms is sometimes elusive in the observations \citep{2022RAA....22g2001T}.

Prevalent oscillations and waves in the corona can be used to infer some parameters including the magnetic field \citep[e.g.,][]{2020ARA&A..58..441N,2020SSRv..216..136L,2021SSRv..217...76B}.
This method was first applied to transverse oscillations of coronal loops \citep[e.g.,][]{Nakariakov2001,2018A&A...617A..86L,2020A&A...638A..32Z} and streamers \citep[e.g.,][]{Chen2011} triggered by flares. These earlier studies can only give average magnetic field strengths along the oscillating structures, and the oscillations often only last for several periods.
It has also been applied to persistent or decayless waves and oscillations in the corona \citep[e.g.,][]{Tomczyk2007,Wang2012,Tian2012} to obtain 2D coronal magnetic field maps \citep{Long2017,Yang2020b,Yang2020a}.
But these studies can only provide the plane-of-sky component of the magnetic field in the off-limb observations.

Another approach is to construct coronal magnetic field models through extrapolations from magnetic field maps in the photosphere \citep[e.g.,][]{Schatten1969,Zhu2018,Wiegelmann2021,2022arXiv220315356Z}.
The combination of magnetic field or magnetohydrodynamic (MHD) models and coronal observations in extreme ultraviolet (EUV) or infrared passbands  can also be used to infer the magnetic field structures in the corona \citep[e.g.,][]{2008ApJ...680.1496L,2009AnGeo..27.2771L,2011ApJ...731L...1D,2016FrASS...3....8G,2018ApJ...856...21C,2019ApJ...883...55Z,2021ApJ...912..141Z,2022RAA....22g5007Z,2022Innov...300236J}.
However, these models are often analytical or based on many assumptions which may not resemble the real Sun \citep[e.g.,][]{Peter2015_extraplation}, {and comprehensive models which can capture both chromospheric and coronal features are needed \citep[e.g.,][]{Aschwanden2016}.
Furthermore, the accuracy of the models is constrained by the limited spatial resolution of the photospheric magnetic field in observations \citep{DeRosa2015}.}
Thus, it is still necessary to directly measure the coronal magnetic field. 

Recently, \citet{Li2015,Li2016} noticed the magnetic-field-induced transition (MIT) in the Fe~{\sc{x}} ion and suggested that it can be used to diagnose coronal magnetic field strength \citep[also see][]{Li2021,2022ApJ...937...48X}.
\citet{Si2020} developed the direct line ratio technique and applied this method to spectral observations taken by EUV Imaging Spectrometer \citep[EIS, ][]{EIS} onboard \textit{Hinode},
and the suitability of the method has been validated through forward modeling with a series of MHD models \citep{Chen2021b,Chen2021,2023RAA....23b2001C,2022ApJ...938....7L,Martinez2022}.
\citet{Landi2020} further developed the weak and strong magnetic field techniques, which has been applied to EIS observations in flare regions \citep{Landi2021} and coronal loops \citep{Brooks2021b,Brooks2021}.
But these techniques have not been validated.
Furthermore, \citet{Chen2021} found that temperature is important for magnetic field measurements and developed a method to simultaneously estimate coronal temperature and density using intensity ratios of the Fe~{\sc{x}} 174/175 and 184/345 {\AA} line pairs to achieve accurate magnetic field measurements. However, the Fe~{\sc{x}} 345 {\AA} is not observed by \textit{Hinode}/EIS or any other instruments in operation.

In this study, we aim to investigate the limitations of coronal magnetic field measurements using EIS observations based on the MIT theory through forward modeling.
We first investigate the suitability of the weak and strong field techniques in \sect{weak_field}. 
Then we propose two methods to diagnose temperature and density in \sect{Temp_diag}, i.e., from the differential emission measure (DEM) analyses and the intensity ratios of the Fe~{\sc{ix}} and Fe~{\sc{xi}} line pairs, respectively.
Finally, we summarize the results in \sect{conclutions}.

%==========================================================================================
\section{Model and Atomic Databases}\label{model}
%==========================================================================================

The radiation MHD model used in this study is the same as the one used in \citet{Chen2021},
and it is calculated using the coronal extension version of the MURaM code \citep{MURaM2005,Rempel2017}. 
The model extends from $\sim$7.5~Mm below to $\sim$41.6~Mm above the photosphere with a grid size of 64~km in the vertical direction.
It contains a region of 98.304$\times$49.152~Mm$^2$ with a grid spacing of 192~km in the horizontal direction. 
%
%Periodic boundary conditions are used in the horizontal direction, and the top boundary is only open to outflows.
%
A torus flux rope is introduced from the bottom boundary and forms an active region containing a bipolar sunspot pair \citep{Rempel2014}.
For more details of the model, we refer the readers to \citet{Rempel2017} and \citet{Chen2021}.

To perform forward modeling with the model, we synthesized the intensities of several coronal emission lines.
The CHIANTI database \citep[version 10.0;][]{CHIANTI,CHIANTIv10} was used to calculate the emissivities.
Because MIT in Fe~{\sc{x}} is not captured by the current version of CHIANTI, we modified the atomic data for the Fe~{\sc{x}} lines, i.e., the radiative transition data is taken from \citet{Wang2020} and the transition probability of MIT is given by \citet{Li2021}.
The Fe~{\sc{x}} lines used in this study are listed in \tab{tab:fexlines}.
The emissivity of the Fe~{\sc{x}} 257.261 {\AA} line consists of two components: one is from the forbidden magnetic quadruple (M2) transition, and the other is from the MIT decay channel.
There is a nearby Fe~{\sc{x}} 257.259 {\AA} line associated with an electric dipole (E1) transition from the 3p$^4$ 3d $^4$D$_{5/2}$ level to the ground state.
As the wavelength difference between the 257.261 {\AA} and 257.259 {\AA} lines is too small to be resolved in observations, we used the total intensity of E1, M2 and MIT lines as the intensity of the 257 {\AA} line.
To derive the formation temperature of the Fe~{\sc{x}} lines using the spectral lines from other ions that can be observed by EIS, we also synthesized the intensities of several lines from the Fe~{\sc{viii}}--{\sc{ix}}, {\sc{xi}}--{\sc{xiii}} ions listed in \tabs{tab:demlines} and \ref{tab:911lines}.
As these lines are not affected by the MIT effect, the original version of the CHIANTI database was used for their calculations.
{It is worth mentioning that the coronal emissions are calculated assuming ionization equilibrium and optically thin radiation and that the absorption of coronal emission from cold plasma is not considered.}

\begin{table}
%\caption{Simulated Fe~{\sc{x}} lines }
\caption{Fe~{\sc{x}} lines used in this study.}
\centering
\begin{tabular}{c|c}
\hline\hline
 Wavelength ({\AA})    & Upper level $\to$ Lower level      \\ \hline
 174.531     & 3s$^2$ 3p$^4$ 3d $^2$D$_{5/2}$ $\to$  3s$^2$ 3p$^5$ $^2$P$_{3/2}$\\ \hline
 175.263    & 3s$^2$ 3p$^4$ 3d $^2$D$_{3/2}$ $\to$  3s$^2$ 3p$^5$ $^2$P$_{1/2}$ \\ \hline
 184.537    & 3s$^2$ 3p$^4$ 3d $^2$S$_{1/2}$ $\to$  3s$^2$ 3p$^5$ $^2$P$_{3/2}$ \\ \hline
 257.259    & 3s$^2$ 3p$^4$ 3d $^4$D$_{5/2}$ $\to$  3s$^2$ 3p$^5$ $^2$P$_{3/2}$ \\ \hline
 257.261    & 3s$^2$ 3p$^4$ 3d $^4$D$_{7/2}$ $\to$  3s$^2$ 3p$^5$ $^2$P$_{3/2}$ \\ \hline
 345.738    & 3s 3p$^6$ $^2$S$_{1/2}$        $\to$  3s$^2$ 3p$^5$ $^2$P$_{3/2}$ \\ \hline
\end{tabular} 
\label{tab:fexlines}
\end{table}

In this study, we assumed that our MHD model is located at the disk center. In other words, a line of sight (LOS) along the vertical direction was chosen. 
To synthesize the intensity maps, we first created lookup tables of the contribution functions G for different lines.
For the Fe~{\sc{x}} lines, the contribution functions $G(T,n_e,B)$ are functions of temperature $T$, electron density $n_e$, and magnetic field strength $B$.
The intensity maps of the Fe~{\sc{x}} lines were taken from \citet{Chen2021}.
For the lines from other ions, the contribution functions $G(T,n_e)$ are functions of temperature and electron density.
The emissivity of each line at each voxel was calculated following $n_e^2G(T,n_e)$ and then integrated along the vertical direction to obtain the intensity map.

%>>>>>>>>>>>>>>>>>>>>>>>>>>>>>>>>>>>>>>>>>>>>>>>>>>>>>>>>>>>>>>>>>>>>>>>>>>>>>>
%\begin{figure} 
%\centering {\includegraphics[width=80mm]{fig_B0.eps}}
%\caption{The magnetic field strength weighted by the Fe~{\sc{x}} 174 {\AA} line emission, reproduced from Figure 3(a) in %\citet{Chen2021}. See \sect{model}.
%} \label{fig_b0}
%\end{figure}
%<<<<<<<<<<<<<<<<<<<<<<<<<<<<<<<<<<<<<<<<<<<<<<<<<<<<<<<<<<<<<<<<<<<<<<<<<<<<<<

%==========================================================================================
\section{Weak and strong magnetic field techniques} \label{weak_field}
%==========================================================================================

We first present a brief introduction to the weak and strong magnetic field techniques.
For a detailed description, we refer the reader to \citet{Landi2020}.
Because the magnetically insensitive E1 transition dominates the blending of the Fe~{\sc{x}} 257 {\AA} line, the intensity variation caused by the MIT effect is not very sensitive to magnetic field strength.
The idea of the weak magnetic field technique is to isolate the intensities from the MIT and M2 transitions, i.e., $I_{MIT}$ and $I_{M2}$, respectively.
Because the MIT and M2 transitions are blended, a reference Fe~{\sc{x}} line in addition to the 257 {\AA} line is needed to eliminate the blend.
When the magnetic field strength is less than 150--200 G, the MIT transition does not change the population of the $^4D_{7/2}$ level much.
Then the changes in the M2 line and other Fe~{\sc{x}} lines are negligible.
The intensity of the MIT transition can be determined as:
\begin{equation}
    I_{MIT}=I_{257}-I_{ref}\times\frac{I_{E1}^{B=0}+I_{M2}^{B=0}}{I_{ref}^{B=0}}
\end{equation}
where $I_{257}$ and $I_{ref}$ are the intensities of the Fe~{\sc{x}} 257 {\AA} line and the reference line, respectively, and $I_i^{B=0}$ ($i=E1, M2, ref$) is the line intensity given by the atomic database neglecting the MIT transition.
Besides, the intensity of the M2 component can be determined as:
\begin{equation}
    I_{M2}=I_{ref}\times\frac{I_{M2}^{B=0}}{I_{ref}^{B=0}}
\end{equation}
Thus the ratio of the two components can be written as:
\begin{equation}
    \frac{I_{MIT}}{I_{M2}}=\frac{I_{257}}{I_{ref}}\times\frac{I_{ref}^{B=0}}{I_{M2}^{B=0}}-\frac{I_{E1}^{B=0}+I_{M2}^{B=0}}{I_{M2}^{B=0}}
    \label{eqn:wfm2}
\end{equation}
here $I_{MIT}/I_{M2}$ is magnetically sensitive and monotonically increases with magnetic field strength (see Figure 1 in \citet{Landi2020}).
After calculating $I_{MIT}/I_{M2}$ using \eqn{eqn:wfm2}, the magnetic field strength can be derived.
When the magnetic field is strong enough to significantly change the population of the $^4D_{7/2}$ level, M2 line intensity will also change with the magnetic field strength, and the weak field technique is no longer valid.
\citet{Landi2020} further developed the strong field technique, in which the influence of the magnetic field strength on the $^4$D$_{7/2}$ population is considered but its influence on other levels is neglected. In other words, only the intensity of the 257.261 {\AA} line is affected by the magnetic field.
Thus the E1 intensity can isolated, and one can obtain:
\begin{equation}
    \frac{I_{MIT}+I_{M2}}{I_{ref}}=\frac{I_{257}}{I_{ref}}-\frac{I_{E1}^{B=0}}{I_{ref}^{B=0}}
    \label{eq:sfm}
\end{equation}
The magnetic field strengths can be derived by comparing $(I_{MIT}+I_{M2})/I_{ref}$ calculated from \eqn{eq:sfm} and the values predicted by theory.
When these two techniques were applied to EIS observations \citep{Landi2020,Landi2021,Brooks2021,Brooks2021b}, they took the 184 {\AA} line as the reference line. $I_{257}$ and $I_{184}$ are the intensities of the 257 and 184 {\AA} lines obtained from observations, and $I_{E1}^{B=0}$, $I_{M2}^{B=0}$, and $I_{184}^{B=0}$ are given by the original version of CHIANTI, which does not account for the MIT transition.
In this paper, the 174 {\AA} line is chosen as the reference line.
It is worth mentioning that the choice of different reference lines among 175, 177, and 184 {\AA} lines will not change our results.
$I_{257}$ and $I_{174}$ were given by synthesized intensity images of the 257 and 174 {\AA} lines, respectively.
After determining the temperature $T^*$ and electron density $n^*$ for each pixel,
the ratios of $I_{174}^{B=0}$, $I_{E1}^{B=0}$, and $I_{M2}^{B=0}$ can be estimated by the ratios of the contribution functions:
\begin{equation}
    \frac{I_{M2}^{B=0}}{I_{174}^{B=0}}=\frac{G_{M2}(T^*,n^*,B=0)}{G_{174}(T^*,n^*,B=0)} \nonumber
\end{equation}
\begin{equation}
    \frac{I_{E1}^{B=0}}{I_{M2}^{B=0}}=\frac{G_{E1}(T^*,n^*,B=0)}{G_{M2}(T^*,n^*,B=0)} \nonumber
\end{equation}
\begin{equation}
    \frac{I_{E1}^{B=0}}{I_{174}^{B=0}}=\frac{G_{E1}(T^*,n^*,B=0)}{G_{174}(T^*,n^*,B=0)} \nonumber
\end{equation}
where $G_{174}$, $G_{E1}$ and $G_{M2}$ are the contribution functions of the Fe~{\sc{x}} 174.531, 257.259 and 257.261 {\AA} lines, respectively.
As the ratios of the contribution functions depend on both the electron density and temperature, we need to determine the density and temperature at each pixel before using \eqn{eqn:wfm2} or (\ref{eq:sfm}) to derive the magnetic field strength.

%>>>>>>>>>>>>>>>>>>>>>>>>>>>>>>>>>>>>>>>>>>>>>>>>>>>>>>>>>>>>>>>>>>>>>>>>>>>>>>
\begin{figure*} 
\centering {\includegraphics[width=\textwidth]{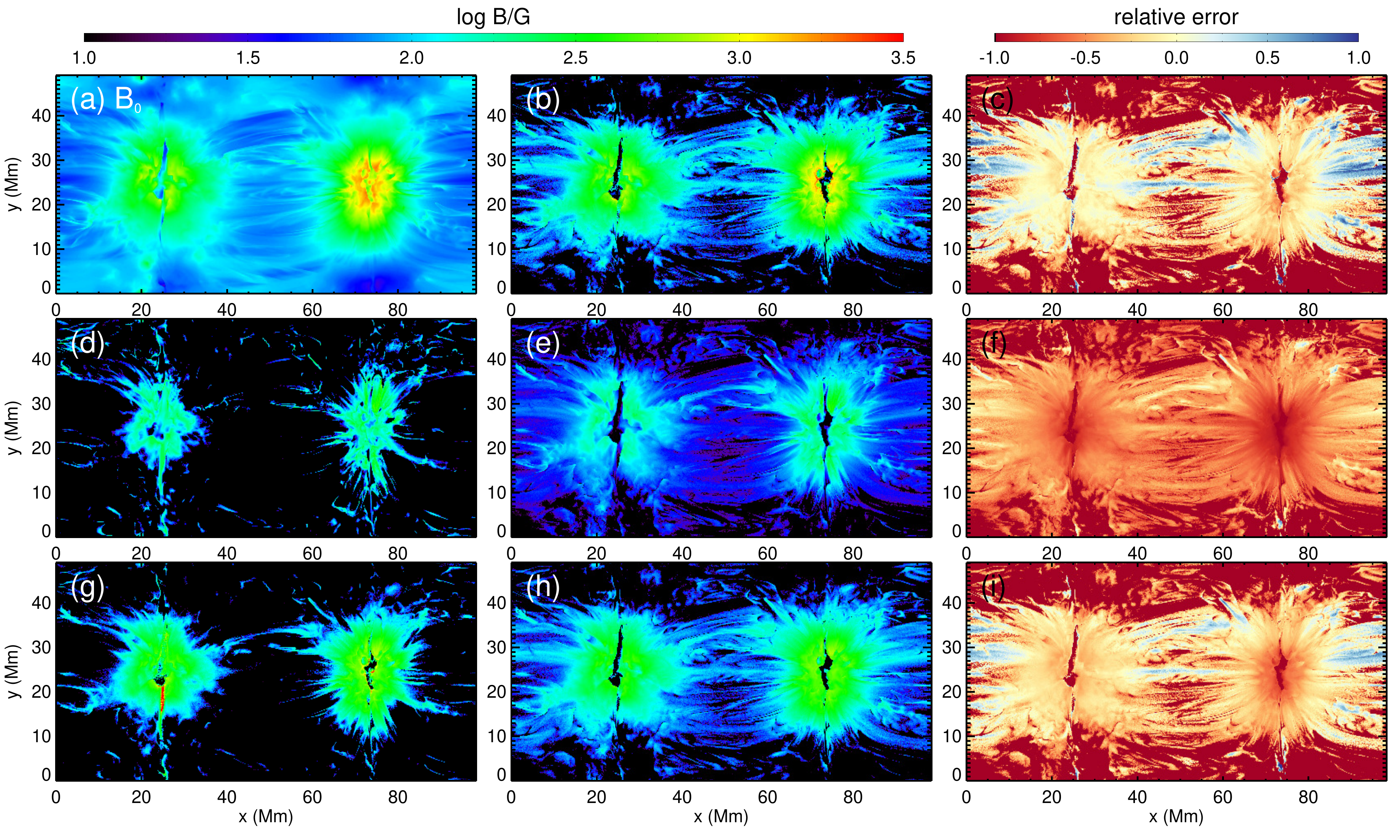}} 
\caption{{The derived coronal magnetic field strength ($B_1$) using the intensity ratios of the Fe~{\sc{x}} 257/174 {\AA} line pair using different techniques and their relative errors.
(a) The magnetic field strength in the model ($B_0$) given by \eqn{eq:b_0}.
(b) $B_1$ calculated from the direct line ratio technique using the density and temperature maps derived from Fe~{\sc{x}} 345/184 and 174/175 {\AA} line pairs. Image reproduced from \citet{Chen2021}.
(c) Relative error ($(B_1-B_0)/B_0$) of the magnetic field map shown in panel (b).
(d) $B_1$ calculated using the weak magnetic field technique. A fixed temperature of $10^{6.0}$ K is assumed for both electric density and magnetic field diagnostics. 
(e) Similar to (d) but using the density and temperature maps derived from Fe~{\sc{x}} 345/184 and 174/175 {\AA} line pairs.
(f) Similar to (c) but for the magnetic field map shown in panel (e).
(g--i) Similar to (d--f) but using the strong magnetic field technique.}
See \sect{weak_field}.
} \label{fig_weak}
\end{figure*}
%<<<<<<<<<<<<<<<<<<<<<<<<<<<<<<<<<<<<<<<<<<<<<<<<<<<<<<<<<<<<<<<<<<<<<<<<<<<<<<

In \citet{Chen2021}, two methods were used to estimate the coronal temperature and density.
The first method assumes a fixed temperature of 10$^{6.0}$ K, where the contribution functions of the Fe~{\sc{x}} lines peak. Then the electron density is derived based on the intensity ratio of the density-sensitive Fe~{\sc{x}} 174/175 {\AA} line pair.
The second method is to simultaneously calculate the electron density and temperature from the Fe~{\sc{x}} 174/175 and 184/345 {\AA} line ratios using a least-squares method.
We took the temperature and density maps obtained from these two methods and derived the magnetic field strengths using the weak field technique from \eqn{eqn:wfm2}. The results are shown in \fig{fig_weak} (d--e), respectively.
Comparing the results to those derived from the direct line ratio technique as shown in Figure 3 in \citet{Chen2021} and \fig{fig_weak} (b) in this paper,
it is obvious that the regions where magnetic field measurements can be performed are roughly the same for the given temperature and density maps.
In other words, the weak field technique cannot improve the suitability of the MIT method.
Following \citet{Chen2021}, we defined the magnetic field strength in the model ($B_0$) as the emission-weighted averaged field strength:
\begin{equation}
    B_0=\frac{\int\epsilon_{174}(z)\cdot B(z)dz}{\int\epsilon_{174}(z)dz}
    \label{eq:b_0}
\end{equation}
where $\epsilon_{174}$ is the emissivity of the Fe~{\sc{x}} 174 {\AA} line given by $n_e^2G_{174}$, {and $B_0$ is presented in \fig{fig_weak} (a) as a reference.
Furthermore, the relative error of the magnetic field map derived from the weak field technique compared to $B_0$ was also calculated and presented in \fig{fig_weak} (f).}
It is obvious that the magnetic field strengths derived using the weak field technique are lower than the values in the model.

Then we calculated the magnetic field maps based on the strong field technique using the same temperature and density maps, and the results are shown in \fig{fig_weak} (g--h).
The strong field technique can only provide magnetic field strengths measurements around the footpoints of coronal loops when assuming a fixed temperature of 10$^{6.0}$ K, which is similar to the weak field technique and direct line ratio technique.
Besides, the magnetic field strengths derived from the strong field technique are mostly larger than the weak field technique, which is consistent with the comparison of weak and strong field techniques in \citet{Brooks2021}.
However, the area coverage of the regions that strong field technique can provide magnetic field strength measurements is larger than the other techniques.
It is because the failure of the direct line ratio technique at the upper parts of coronal loops results from the temperature underestimation \citep{Chen2021}, and the temperature dependence of $(I_{MIT}+I_{M2})/I_{174}$ is less than that of $I_{257}/I_{174}$.
When more accurate density and temperature maps are used, the suitability of the strong field technique are almost the same as the direct line ratio technique.
{The relative error of the calculated magnetic field strength using the strong field technique is also shown in \fig{fig_weak} (i).}
The strong field technique underestimates coronal magnetic field strengths, especially for the regions with strong magnetic field strengths, i.e., sunspot regions.

%>>>>>>>>>>>>>>>>>>>>>>>>>>>>>>>>>>>>>>>>>>>>>>>>>>>>>>>>>>>>>>>>>>>>>>>>>>>>>>
\begin{figure*} 
\centering {\includegraphics[width=\textwidth]{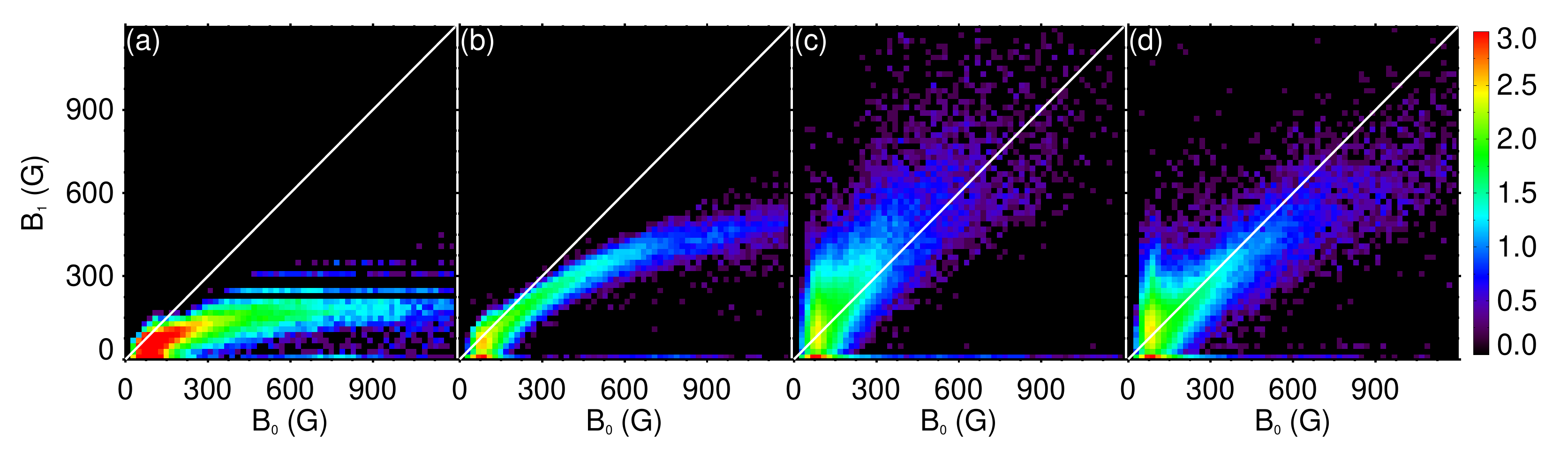}} 
\caption{Joint PDFs of $B_0$ and the MIT-measured field strength ($B_1$) derived from different techniques. (a-b) corresponds to the results derived from weak and strong magnetic field techniques (\fig{fig_weak} (e,h)), respectively. (c-d) similar to (a) but using direct line ratio technique based on the temperature diagnostics with DEM analysis (\fig{fig_dem2} (c)) and Fe~{\sc{ix}} and Fe~{\sc{xi}} line ratios (\fig{fig_911} (a)), respectively. 
} \label{fig_pdf}
\end{figure*}
%<<<<<<<<<<<<<<<<<<<<<<<<<<<<<<<<<<<<<<<<<<<<<<<<<<<<<<<<<<<<<<<<<<<<<<<<<<<<<<

To better compare the results and understand the limitations of the weak and strong field techniques, we also calculated the joint probability density functions (PDF) of $B_0$ and magnetic field strengths derived using the two MIT techniques ($B_1$)  as shown in \fig{fig_pdf}(a--b).
The solid white lines in \fig{fig_pdf} indicate that the MIT-measured magnetic field strengths are the same as the values in the model.
The magnetic field strengths derived from the weak magnetic field technique are significantly lower than the values in the model, and there is a weak correlation between $B_0$ and $B_1$.
The results of the strong field technique is more reasonable than the weak field technique, i.e., there is a clear correlation between $B_0$ and $B_1$.
However, the joint PDF of the strong field technique starts to deviate from $B_0=B_1$ when magnetic field strength is larger than $\sim$300 G, while direct line ratio technique slightly underestimates field strengths above $\sim$600 G \citep{Chen2021}.
The limitation of the weak and strong magnetic field techniques are caused by inconsistency in the theory.
Even if the external magnetic field strength is weak, the population of the levels in the Fe~{\sc{x}} ion will change.
But the weak and strong field techniques (partially) neglect the changes in the level population induced by the MIT effect, which results in systematic deviation in measured magnetic field strength.
The deviation is more prominent when the magnetic field strength is strong.
As neither the weak and strong field techniques can provide as accurate magnetic field measurements as the direct line ratio technique, we choose the direct line ratio technique to derive the magnetic field strength in the following section.

%==========================================================================================
\section{Temperature determination using lines available in EIS observations}\label{Temp_diag}
%==========================================================================================

\citet{Chen2021} found that temperature diagnostics is essential for accurate magnetic field measurements based on the MIT theory, and they introduced a temperature-sensitive Fe~{\sc{x}} 184/345 {\AA} line pair to diagnose the temperature.
Nevertheless, the 345 {\AA} line is not within the wavelength ranges of the EIS detectors, so their temperature diagnosing method cannot be applied to available spectroscopic observations.
Thus, we need to explore other approaches to estimate the formation temperature of the Fe~{\sc{x}} lines {from spectral lines available in EIS observations}.
In this section, we focus on the accuracy of temperature determination from DEM analysis in \sect{DEM} and Fe~{\sc{ix}} and {\sc{xi}} line ratios in \sect{Fe_911}.

%------------------------------------------------------------------------------------------
\subsection{Temperature diagnostics based on DEM analysis}\label{DEM}
%------------------------------------------------------------------------------------------

%>>>>>>>>>>>>>>>>>>>>>>>>>>>>>>>>>>>>>>>>>>>>>>>>>>>>>>>>>>>>>>>>>>>>>>>>>>>>>>
\begin{table}
%\caption{Simulated Fe~{\sc{x}} lines }
\caption{Spectral lines used for the DEM analysis.}
\centering
\begin{tabular}{c|c|c}
\hline\hline
 Ion name & Wavelength ({\AA})    & log$_{10}$ T$_{max}$/K      \\ \hline
 Fe~{\sc{viii}} & 185.213  & 5.65 \\ \hline
 Fe~{\sc{ix}}   & 185.493  & 5.90 \\ \hline
 Fe~{\sc{x}}    & 184.537  & 6.00 \\ \hline
 Fe~{\sc{xi}}   & 188.216  & 6.10 \\ \hline
 Fe~{\sc{xii}}  & 195.119  & 6.20 \\ \hline
 Fe~{\sc{xiii}} & 202.044  & 6.25 \\ \hline
\end{tabular} 
\label{tab:demlines}
\end{table}
%<<<<<<<<<<<<<<<<<<<<<<<<<<<<<<<<<<<<<<<<<<<<<<<<<<<<<<<<<<<<<<<<<<<<<<<<<<<<<<

Although there is no suitable temperature-sensitive Fe~{\sc{x}} line pair within the wavelength range of EIS, we can diagnose the temperature using spectral lines from other Fe ions. One of the most popular methods is DEM analysis using spectral observations of a number of lines coving a wide temperature range \citep{Del_Zanna2018}.
Many DEM methods have been developed over the past few decades \citep[e.g.,][]{Kashyap1998,Landi2002,aschwanden2019new}, and we chose the regularized inversion technique developed by \citet{Hannah2012} to derive the temperature distribution at each pixel.
The spectral lines used for the DEM analysis are listed in \tab{tab:demlines}, and these lines provide good temperature coverage around 10$^{6.0}$ K.
{The temperature range used for the DEM inversion is 10$^{5.4}$ to 10$^{6.6}$ K with a temperature bin size of $\Delta({\mathrm{log_{10}}}~T/{\mathrm{K}})=0.05$.}
The intensity maps of these lines are synthesized as described in \sect{model}.

%>>>>>>>>>>>>>>>>>>>>>>>>>>>>>>>>>>>>>>>>>>>>>>>>>>>>>>>>>>>>>>>>>>>>>>>>>>>>>>
\begin{figure*} 
\centering {\includegraphics[width=13cm]{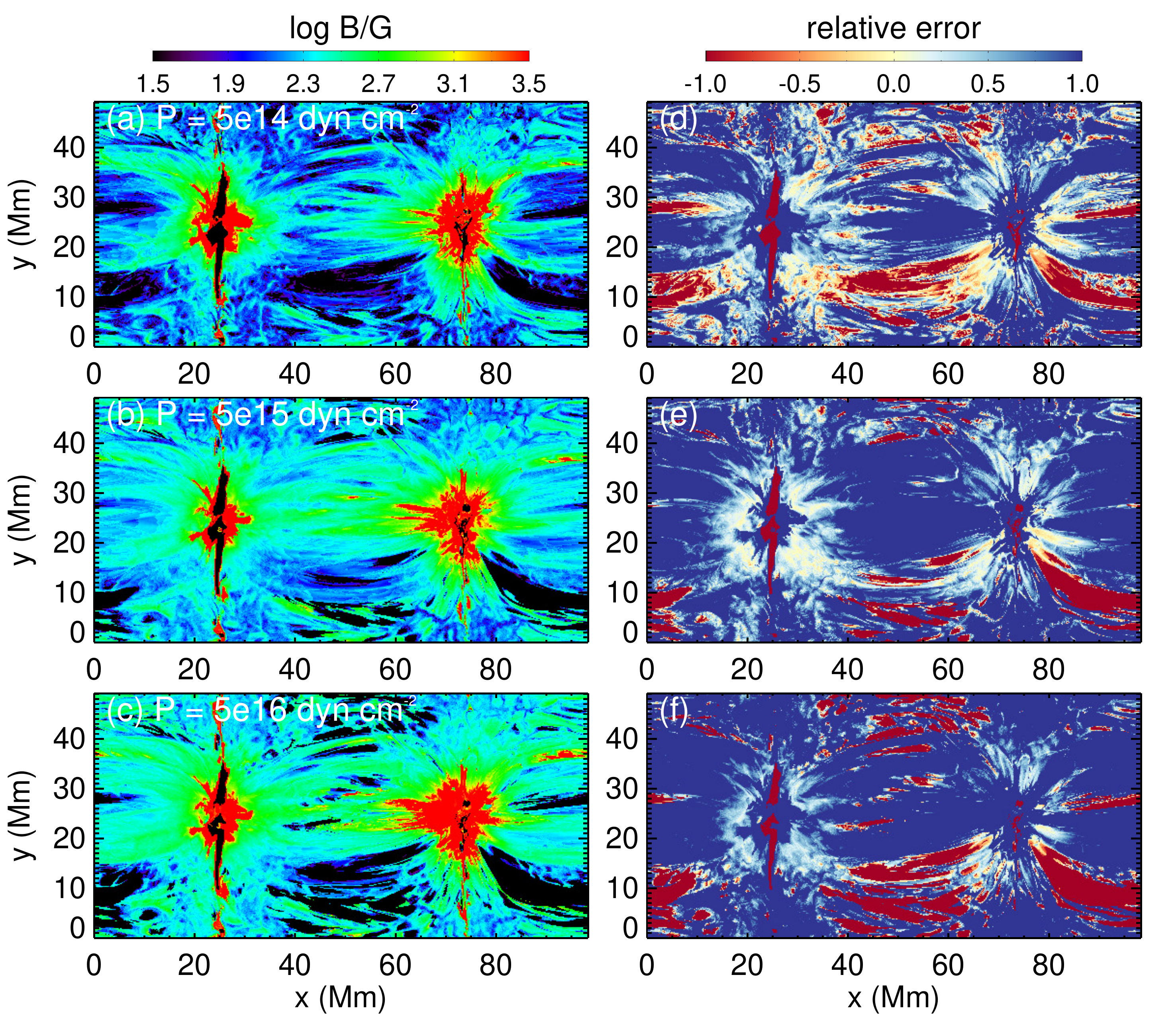}}
\caption{The derived magnetic field strength using the Fe~{\sc{x}} 257/174 {\AA} line pair.
(a--c) The pressure used for DEM analysis is assumed to be $5\times10^{14}$, $5\times10^{15}$, and $5\times10^{16}$ dyn cm$^{-2}$, respectively.
The temperature used for density and magnetic field strength estimations is given by \eqn{temp}.
Note that the dynamic range of the magnetic field strength in this figure is different from that in other figures. 
{(d--f) Relative errors of the magnetic field maps shown in panels (a)--(c)}.
See \sect{DEM}.
} \label{fig_dem1}
\end{figure*}
%<<<<<<<<<<<<<<<<<<<<<<<<<<<<<<<<<<<<<<<<<<<<<<<<<<<<<<<<<<<<<<<<<<<<<<<<<<<<<<

\begin{figure} 
\centering {\includegraphics[width=80mm]{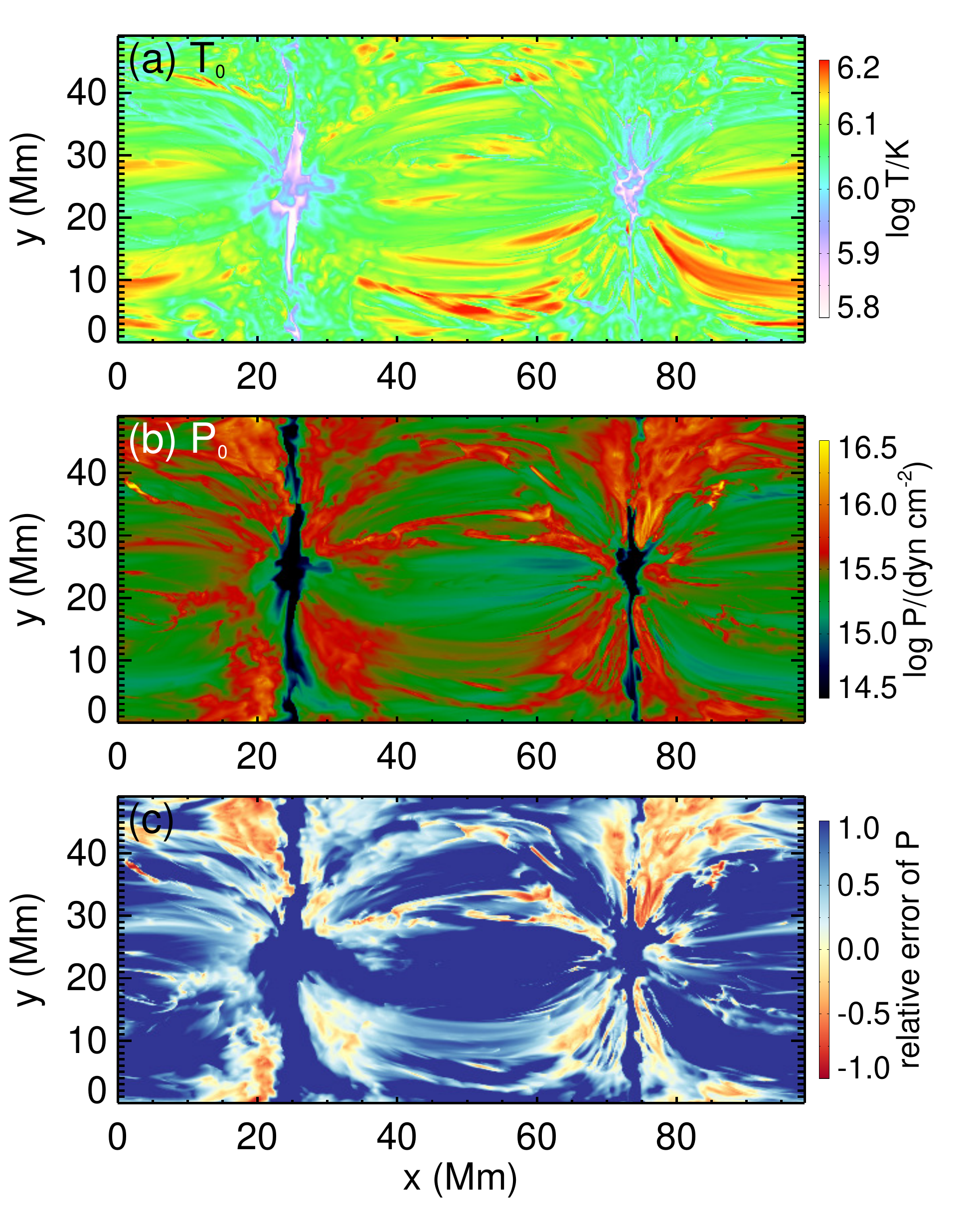}}
\caption{{(a--b) Emission-weighted average temperature and pressure given by \eqns{eqn:t_0} and (\ref{eqn:p_0}), respectively.
(c) Relative error of the pressure ($(P_1-P_0)/P_0$, where $P_1=5\times10^{15}$ dyn cm$^{-2}$).
}
See \sect{DEM}.
} \label{fig_t0_p0}
\end{figure}

When performing the DEM analysis, a constant pressure is required to ensure the contribution functions of the spectral lines as functions of the temperature only.
It is worth noting that the temperature sampling of the DEM results is uniformly distributed in the logarithmic scale. 
For each pixel, the line intensity can be expressed as:
\begin{equation}
    I=\int G(T)\cdot {\rm{DEM}}(T)~dT={\rm{ln}}10\int G(T)\cdot {\rm{DEM}}(T)\cdot T~d({\rm{log}}_{10}T)
\end{equation}
After obtaining the DEM(T) from the regularized inversion technique, we determined an emission-weighted averaged temperature at each pixel as:
\begin{equation}
\begin{aligned}
    T_X^*
    &=\frac{\int G_{174}(T)\cdot{\mathrm{DEM}}(T)\cdot T~d(T)}{\int G_{174}(T)\cdot{\mathrm{DEM}}(T)~d(T)}\\
    &=\frac{\int G_{174}(T)\cdot{\mathrm{DEM}}(T)\cdot T\cdot T~d({\rm{log}}_{10}T)}{\int G_{174}(T)\cdot{\mathrm{DEM}}(T)\cdot T~d({\rm{log}}_{10}T)}  \label{temp}
\end{aligned}
\end{equation}
Furthermore, we assumed a temperature of $T_X^*$ and derive the electron density $n_X^*$ from the Fe~{\sc{x}} 174/175 {\AA} line ratio.
After that, the magnetic field strength can be calculated using the Fe~{\sc{x}} 257/174 {\AA} line ratios. 
We took three pressure values of $5\times10^{14}$, $5\times10^{15}$, and $5\times10^{16}$ dyn~cm$^{-2}$ for DEM analysis, respectively, and the corresponding MIT-measured magnetic field strengths are shown in \fig{fig_dem1} (a--c).
{Their relative errors compared to $B_0$ are also presented in \fig{fig_dem1} (d--e).}
It is obvious that the MIT-measured magnetic field strength depends on the choice of the pressure used for DEM analysis.
Actually, the contribution functions depend not only on temperature but also pressure, but we did not account for dependence on pressure for the simplification of the DEM inversion. 
{Similar to \eqn{eq:b_0}, we derived the electron pressure in the model ($P_0$) as the emission-weighted average values:}
\begin{equation}
    P_0=\frac{\int\epsilon_{174}(z)\cdot P(z)dz}{\int\epsilon_{174}(z)dz}
    \label{eqn:p_0}
\end{equation}
{and $P_0$ is presented in \fig{fig_t0_p0} (b).
It is evident that $P_0$ is highly non-uniform.
We also calculated the relative error of the constant pressure of $5\times10^{15}$ dyn~cm$^{-2}$ compared to $P_0$, and the results are shown in \fig{fig_t0_p0} (c).
The uncertainty of the pressure used for DEM inversion is considerable,}
and the choice of pressure will significantly impact the DEM results and also the calculated magnetic field strengths.

%>>>>>>>>>>>>>>>>>>>>>>>>>>>>>>>>>>>>>>>>>>>>>>>>>>>>>>>>>>>>>>>>>>>>>>>>>>>>>>
%<<<<<<<<<<<<<<<<<<<<<<<<<<<<<<<<<<<<<<<<<<<<<<<<<<<<<<<<<<<<<<<<<<<<<<<<<<<<<<

%>>>>>>>>>>>>>>>>>>>>>>>>>>>>>>>>>>>>>>>>>>>>>>>>>>>>>>>>>>>>>>>>>>>>>>>>>>>>>>
\begin{figure*} 
\centering {\includegraphics[width=\textwidth]{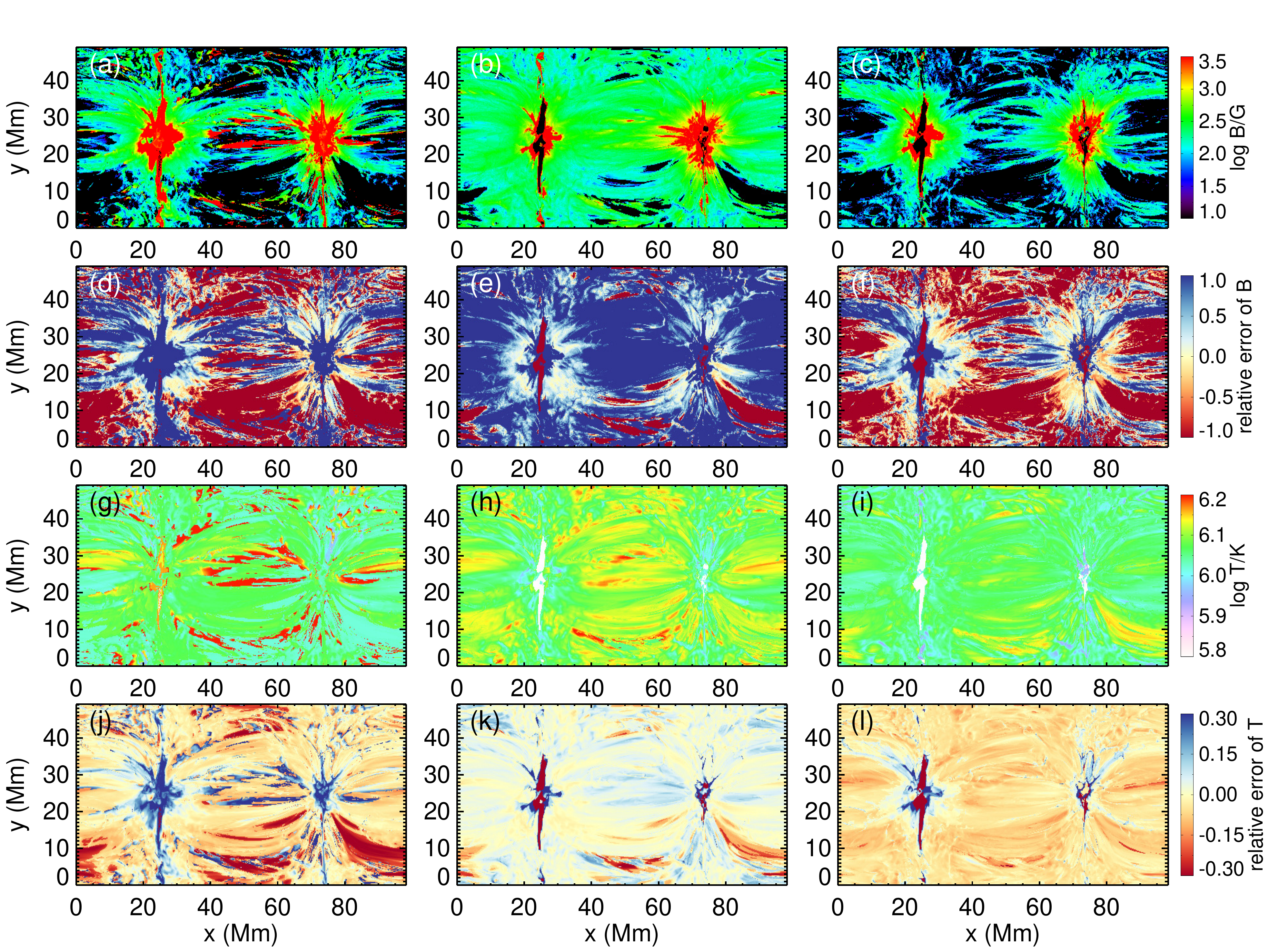}}
\caption{(a--c) The derived magnetic field strength using the Fe~{\sc{x}} 257/174 {\AA} line pair.
The temperature used for the diagnostics of electron density and magnetic field strength are given by $T_{X1}^*$, $T_{X}^*$, and $T_{X2}^*$ {and shown in panels (g--i)}, respectively.
We assume a fixed pressure of $5\times10^{15}$ dyn cm$^{-2}$ for DEM analyses.
{(d--f) Relative error of magnetic field maps shown in panels (a--c), respectively.
(j--l) Relative errors of temperature maps shown in panels (g--i) compared to $T_0$ shown in \fig{fig_t0_p0}(a), respectively.
}
See \sect{DEM}.
} \label{fig_dem2}
\end{figure*}
%<<<<<<<<<<<<<<<<<<<<<<<<<<<<<<<<<<<<<<<<<<<<<<<<<<<<<<<<<<<<<<<<<<<<<<<<<<<<<<

%>>>>>>>>>>>>>>>>>>>>>>>>>>>>>>>>>>>>>>>>>>>>>>>>>>>>>>>>>>>>>>>>>>>>>>>>>>>>>>
\begin{figure*} 
\centering {\includegraphics[width=\textwidth]{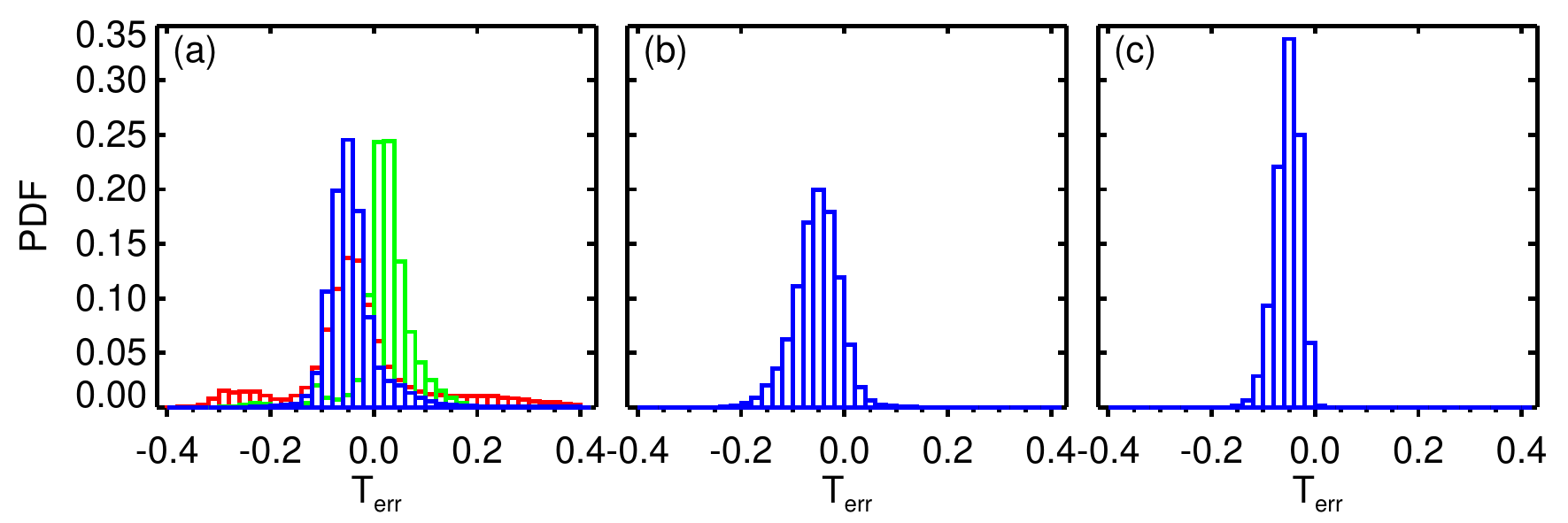}} 
\caption{The histogram of T$_{err}$.
(a) The red, green, and blue histograms are the results given by $T_{X1}^*$, $T_{X}^*$, and $T_{X2}^*$, respectively.
(b) The results obtained from temperature map derived from Fe~{\sc{ix}} and Fe~{\sc{xi}} line pairs.
(c) The same as panel (b) but for temperature calculated using Fe~{\sc{x}} 184/345 and 174/175 {\AA} line pairs.
} \label{fig_terr}
\end{figure*}
%<<<<<<<<<<<<<<<<<<<<<<<<<<<<<<<<<<<<<<<<<<<<<<<<<<<<<<<<<<<<<<<<<<<<<<<<<<<<<<

Besides, we also found that the calculated magnetic field strengths are quite different from $B_0$ no matter which pressure is used for DEM analysis.
To improve the results, we proposed two other approaches to estimate the temperature from the DEM results.
One method is to take the temperature $T_{X1}^*$ where $G_{174}(T)\cdot {\rm{DEM}}(T)\cdot T$ peaks.
It corresponds to the temperature where the Fe~{\sc{x}} 174 {\AA} emission peaks along the LOS.
The other method is to take harmonic weighted averaged temperature as:
\begin{equation}
\begin{aligned}
    T_{X2}^*
    &=\frac{\int G_{174}(T)\cdot{\mathrm{DEM}}(T)~d(T)}{\int G_{174}(T)\cdot{\mathrm{DEM}}(T)\cdot \frac{1}{T}~d(T)}\\
    &=\frac{\int G_{174}(T)\cdot{\mathrm{DEM}}(T)\cdot T~d({\rm{log}}_{10}T)}{\int G_{174}(T)\cdot{\mathrm{DEM}}(T)~d({\rm{log}}_{10}T)}  \label{temp_2}
\end{aligned}
\end{equation}
We used the DEM results with a fixed pressure of $5\times10^{15}$ dyn cm$^{-2}$ to derive $T_{X1}^*$ and $T_{X2}^*$ at each pixel, respectively.
In order to evaluate the accuracy of temperature measurements, we defined the temperature difference ($T_{err}$) between the measured temperature ($T^*$) and the values in the model ($T_0$) as:
\begin{equation}
    T_{err}=(T^*-T_0)/T_0
\end{equation}
where $T_0$ is emission-weighted averaged temperature:
\begin{equation}
    T_0=\frac{\int\epsilon_{174}(z)\cdot T(z)dz}{\int\epsilon_{174}(z)dz}
    \label{eqn:t_0}
\end{equation}
{and $T_0$ is shown in \fig{fig_t0_p0} (a).}
{The temperature and $T_{err}$ maps given by $T_{X1}^*$, $T_{X}^*$, and $T_{X2}^*$ are presented in \fig{fig_dem2} (g--l).}
\fig{fig_terr} (a) presents the histograms of $T_{err}$ corresponding to $T_{X1}^*$, $T_{X}^*$, and $T_{X2}^*$, and their widths are larger than that obtained from the Fe~{\sc{x}} 174/175 and 184/345 {\AA} line pairs as shown in panel (c).
In other words, the temperature measurements based on the DEM analyses from spectral lines of different ions are worse than the least-squares method based on several Fe~{\sc{x}} lines.
Then we calculated the electron density and magnetic field strength from intensity ratios of the 174/175 and 257/174 {\AA} line pairs based on the temperature maps given by $T_{X1}^*$, $T_{X}^*$, and $T_{X2}^*$, respectively.
The derived magnetic field maps are shown in \fig{fig_dem2} (a--c), and they always show evident discrepancies compared to $B_0$, {as illustrated by relative errors of the magnetic field shown in \fig{fig_dem2} (d--f)}.
Temperature determination using \eqn{temp_2} seems to provide the best coronal magnetic field measurements based on the DEM analysis (as shown in \fig{fig_dem2}(c)), and the joint PDF of the MIT-measured field strength and $B_0$ is presented in \fig{fig_pdf}(c).
The correlation is weak, indicating the temperature determination based on DEM analysis can hardly provide reasonable magnetic field measurements. 
Actually, the spatial distribution of the pressure is not uniform in the corona. 
Thus, a uniform pressure used for DEM analysis will result in errors in the derived temperature,
and hence the errors of the derived magnetic field strength are magnified.
{It is worth mentioning that the DEM technique described in \citet{Martinez2022}, developed based on the L1 norm inversion methods in \citet{DEM_Cheung}, can take into account temperature, density, and magnetic field simultaneously, which might improve the accuracy of the temperature and density diagnostics.}

%------------------------------------------------------------------------------------------
\subsection{Temperature and density diagnostics using Fe~{\sc{ix}} and Fe~{\sc{xi}} line ratios}\label{Fe_911}
%------------------------------------------------------------------------------------------

%>>>>>>>>>>>>>>>>>>>>>>>>>>>>>>>>>>>>>>>>>>>>>>>>>>>>>>>>>>>>>>>>>>>>>>>>>>>>>>
\begin{table}
\caption{Fe~{\sc{ix}} and {\sc{xi}} lines used in this study}
\centering
\begin{tabular}{c|c|c}
\hline\hline
Ion name & Wavelength ({\AA}) & log$_{10}$ T$_{max}$/K      \\ \hline
Fe~{\sc{ix}} & 171.073 & 5.90 \\ 
             & 188.493 &      \\ \hline
Fe~{\sc{xi}} & 182.167 & 6.10 \\ 
             & 188.216 &      \\ 
             & 257.547 & \\
             & 257.554 & \\ \hline
\end{tabular} 
\label{tab:911lines}
\end{table}
%<<<<<<<<<<<<<<<<<<<<<<<<<<<<<<<<<<<<<<<<<<<<<<<<<<<<<<<<<<<<<<<<<<<<<<<<<<<<<<

The Fe~{\sc{ix}} 171.07/188.49 {\AA} and Fe~{\sc{xi}} 188.22/257.55 {\AA} line pairs are temperature sensitive \citep{Del_Zanna2018}, and these spectral lines can be observed by EIS simultaneously \citep[e.g.,][]{Young2007}.
It is worth noting that the Fe~{\sc{xi}} 257.55 {\AA} line is self-blended in EIS observations, as the wavelength difference between the 257.547 and 257.554 {\AA} lines is too small to be resolved by EIS. Thus the intensity of the Fe~{\sc{xi}} 257.55 {\AA} line is the total intensity of the 257.547 and 257.554 {\AA} lines.
As the formation temperature of the Fe~{\sc{x}} lines is in between those of the Fe~{\sc{ix}} and Fe~{\sc{xi}} lines, we can calculate the formation temperatures of the Fe~{\sc{ix}} and Fe~{\sc{xi}} lines, respectively, and then take their average.
By examining the CHIANTI atomic database, we found that the Fe~{\sc{ix}} 171.07/188.49 {\AA} line ratio is only sensitive to the temperature and almost does not change with the electron density.
However, the Fe~{\sc{xi}} 188.22/257.55 {\AA} line ratio changes with both the temperature and density.
Thus we also included the density-sensitive Fe~{\sc{xi}} 188.22/182.17 {\AA} line pair to estimate the electron density.
The Fe~{\sc{ix}} and Fe~{\sc{xi}} lines used in this study are summarized in \tab{tab:911lines}.

We first used the Fe~{\sc{ix}} 171.07/188.49 {\AA} line ratio to derive the formation temperature of the Fe~{\sc{ix}} lines ($T_{IX}$).
Then we applied the least squares method developed by \citet{Chen2021} for temperature and density estimations to determine the formation temperature and electron density of the Fe~{\sc{xi}} lines simultaneously.
We defined a norm value ($L$) as:
\begin{equation}
L=\left(\frac{G_{188}^{X1}(T,n_e)}{G_{257}^{X1}(T,n_e)}-\frac{I_{188}^*}{I_{257}^*}  \right)^2-\left(\frac{G_{188}^{X1}(T,n_e)}{G_{182}^{X1}(T,n_e)}-\frac{I_{188}^*}{I_{182}^*}  \right)^2
\end{equation}
where $I_{182}^*$, $I_{188}^*$, and $I_{257}^*$ are the intensities of the Fe~{\sc{xi}} 182.17, 188.22, and 257.55 {\AA} lines at each pixel, respectively, and $G_{182}^{X1}$, $G_{188}^{X1}$, and $G_{257}^{X1}$ are the contribution functions of the three lines.
The temperature ($T_{XI}$) and density ($n_{XI}$) of the Fe~{\sc{xi}} lines are simultaneously determined when $L$ reaches the minimum.

%>>>>>>>>>>>>>>>>>>>>>>>>>>>>>>>>>>>>>>>>>>>>>>>>>>>>>>>>>>>>>>>>>>>>>>>>>>>>>>
\begin{figure*} 
\centering {\includegraphics[width=150mm]{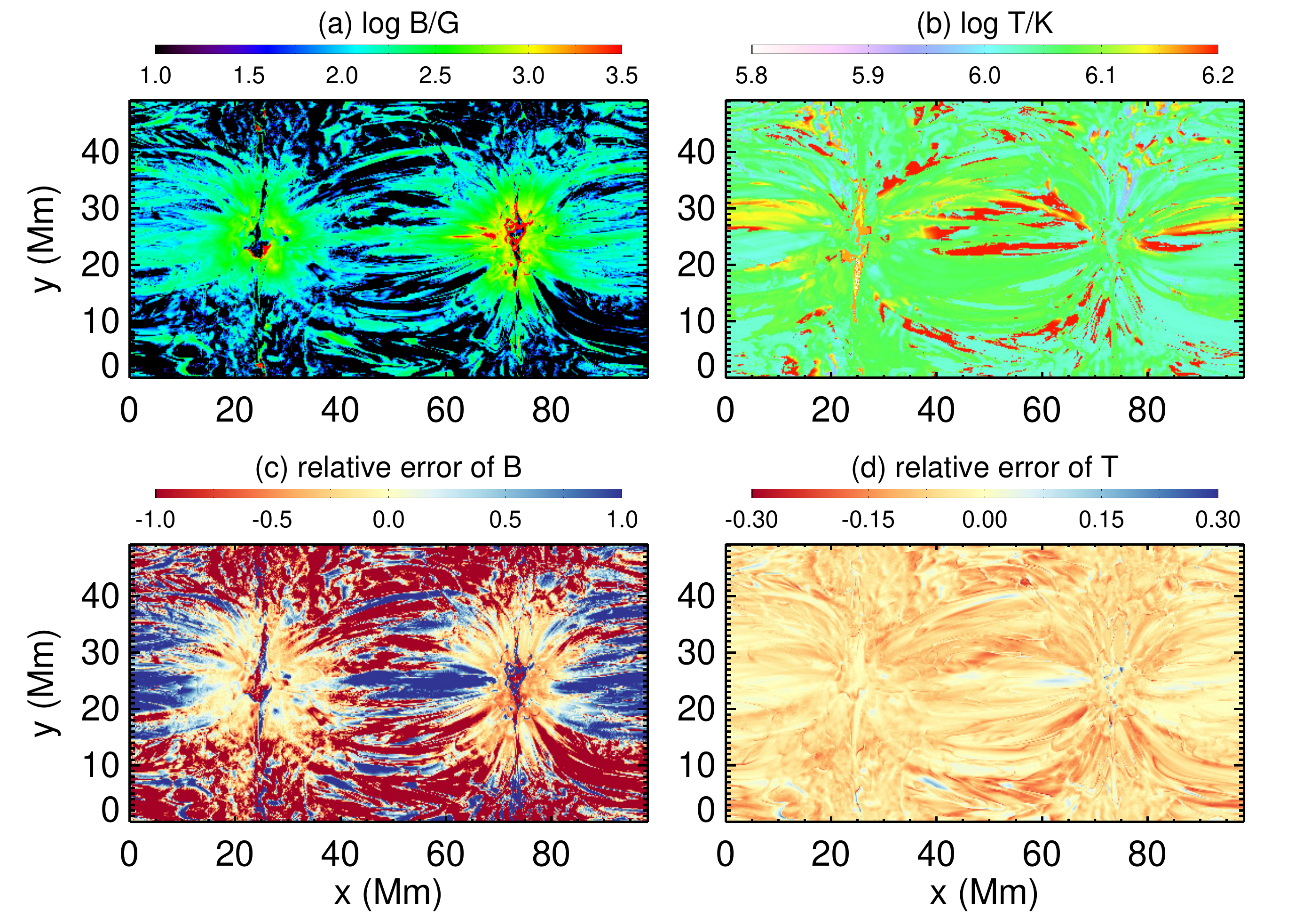}}
\caption{(a) The derived magnetic field strength using the Fe~{\sc{x}} 257/174 {\AA} line pair.
{The temperature of the Fe~{\sc{x}} line estimated from intensity ratios of the Fe~{\sc{ix}} and Fe~{\sc{xi}} line pairs is shown in panel (b).
(c--d) Relative errors of the magnetic field and temperature shown in panels (a--b), respectively.
}
See \sect{Fe_911}.
} \label{fig_911}
\end{figure*}
%<<<<<<<<<<<<<<<<<<<<<<<<<<<<<<<<<<<<<<<<<<<<<<<<<<<<<<<<<<<<<<<<<<<<<<<<<<<<<<

When the temperature of the Fe~{\sc{ix}} and Fe {\sc{xi}} lines are estimated, we took their average as the formation temperature of the Fe~{\sc{x}} lines ($T^*_{2}$).
Then we used the Fe~{\sc{x}} 174/175 {\AA} line ratios to derive the electron density.
Based on the derived temperature and density, the magnetic field strengths were calculated from the Fe~{\sc{x}} 257/174 {\AA} line ratio.
{The calculated temperature and magnetic field strength are shown in \fig{fig_911}, and their relative errors are also presented.}
Compared to the MIT-measured field strengths based on temperature diagnostics using the DEM analysis, this method improves the accuracy of the magnetic field measurements.
Especially for the regions around the sunspots, the magnetic field strength shown in \fig{fig_911} is not highly overestimated compared to those presented in \figs{fig_dem1} and \ref{fig_dem2}.
However, the field strengths around loop apex regions are quite different from the values in the model. 
We present differences between $T^*_{2}$ and $T_{0}$ in \fig{fig_terr}(b), and the accuracy of temperature measurements is not significantly improved compared to those from DEM analyses. 
Compared to the technique proposed in \citet{Chen2021}, the larger uncertainty of temperature determination using Fe~{\sc{ix}} and Fe~{\sc{xi}} line ratios results in larger errors in coronal magnetic field measurements.
The temperature uncertainty is large because the quantitative relationship among the formation temperatures of the Fe~{\sc{ix}}, Fe~{\sc{x}}, and Fe~{\sc{xi}} lines is not clear.
To better evaluate the accuracy of this technique, we also exhibit the joint PDF of the derived magnetic field strength and $B_0$ in \fig{fig_pdf}(d).
Considering the obvious deviation of the joint PDF from the $B_0=B_1$ line in \fig{fig_pdf}(d), this method may not provide an accurate magnetic field measurements in real observations.

%==========================================================================================
\section{Conclusions} \label{conclutions}
%==========================================================================================

In this study, we performed forward modeling with a 3D radiation MHD model of an active region to investigate the limitations of applying the MIT method to EIS observations to obtain coronal magnetic field strength.
{We first synthesized intensities of different coronal emission lines under assumptions of ionization equilibrium, optically thin radiation, and no absorption from cold plasma along the LOS.}
Then we applied the weak and strong field techniques developed by \citet{Landi2020} to obtain magnetic field strengths and then compared the results to that derived from the direct line ratio technique and the values in the model.
We found that the weak field technique cannot improve the suitability of the MIT method but systematically underestimate the coronal magnetic field strength.
The strong field technique can provide more accurate magnetic field strength compared to the weak field technique but also underestimate field strength.
It is because the weak and strong field technique (partially) assumes that the population of Fe~{\sc{x}} levels does not change with magnetic field strength, which is not consistent with the reality.

Furthermore, we applied two temperature diagnosing methods based on the spectral lines within the wavelength range of EIS detectors.
The first method is to perform DEM analysis using the lines listed in \tab{tab:demlines} and then determine the temperature from the DEM distribution.
The second method is to derive the temperature of the Fe~{\sc{ix}} and Fe~{\sc{xi}} lines using the lines listed in \tab{tab:911lines} and then take their average.
Unfortunately, the temperature maps derived from the two methods can not provide as accurate coronal magnetic field measurements as those derived from the Fe~{\sc{x}} 174/175 and 184/345 {\AA} line pairs.
Thus the Fe~{\sc{x}} 345 {\AA} line seems necessary for reasonably accurate temperature and magnetic field measurements. 
However, the 345 {\AA} line cannot be captured by the EIS detectors, and we still need to find a better strategy to achieve accurate magnetic field measurements from EIS observations based on the MIT theory.

\section*{Acknowledgements}

This work is supported by the National Key R\&D Program of China No. 2021YFA0718600 and NSFC grants 11825301, 12103066.

%%%%%%%%%%%%%%%%%%%%%%%%%%%%%%%%%%%%%%%%%%%%%%%%%%
\section*{Data Availability}
The data underlying this article will be shared on reasonable request to the corresponding author.

%The inclusion of a Data Availability Statement is a requirement for articles published in MNRAS. Data Availability Statements provide a standardised format for readers to understand the availability of data underlying the research results described in the article. The statement may refer to original data generated in the course of the study or to third-party data analysed in the article. The statement should describe and provide means of access, where possible, by linking to the data or providing the required accession numbers for the relevant databases or DOIs.

%%%%%%%%%%%%%%%%%%%% REFERENCES %%%%%%%%%%%%%%%%%%

% The best way to enter references is to use BibTeX:

\bibliographystyle{mnras}
\bibliography{ms} % if your bibtex file is called example.bib

% Alternatively you could enter them by hand, like this:
% This method is tedious and prone to error if you have lots of references
%\begin{thebibliography}{99}
%\bibitem[\protect\citeauthoryear{Author}{2012}]{Author2012}
%Author A.~N., 2013, Journal of Improbable Astronomy, 1, 1
%\bibitem[\protect\citeauthoryear{Others}{2013}]{Others2013}
%Others S., 2012, Journal of Interesting Stuff, 17, 198
%\end{thebibliography}

%%%%%%%%%%%%%%%%%%%%%%%%%%%%%%%%%%%%%%%%%%%%%%%%%%

% Don't change these lines
\bsp	% typesetting comment
\label{lastpage}
\end{document}